\documentclass[prl,twocolumn]{revtex4-2}

\usepackage{}
\usepackage{epsfig}
\usepackage{graphicx}
\usepackage{amsfonts,amssymb}
\usepackage{amsmath}
\usepackage{color}
\usepackage {times}
\usepackage{mathrsfs}
\definecolor{refcolor}{rgb}{1.0,0.0,0.0}

\begin{document}

\title{Tunneling maps, non-monotonic resistivity, and non Drude optics
in EuB$_6$}

\author{Tanmoy Mondal and Pinaki Majumdar}

\affiliation{Harish-Chandra Research Institute 
(A CI of Homi Bhabha National Institute), 
Chhatnag Road, Jhusi, Allahabad 211019}
\date{\today}

\begin{abstract}
For several decades the low carrier density local moment magnet EuB$_6$ has 
been considered a candidate material for ferromagnetic polarons.  There is 
however no consistent explanation for the host of intriguing observations 
that have accrued over the years, including a prominently non-monotonic 
resistivity near $T_c$, and observation of spatial textures, with a 
characteristic spatial and energy scale, via scanning tunneling spectroscopy. 
We resolve all these features using a Heisenberg-Kondo 
lattice model for EuB$_6$, solved using exact diagonalisation based Langevin 
dynamics. Over a temperature window $\sim 0.7T_c - 1.5T_c$ 
we observe electronic and magnetic textures with the correct spatial and energy 
scale, and confirm an associated non-monotonic resistivity.  We predict a 
distinctly `non Drude' optical conductivity in the polaronic phase, and propose 
a field-temperature phase diagram testable through spin resolved tunneling 
spectroscopy. We argue that the anomalous properties of EuB$_6$, and magnetic 
polaron materials in general, occur due to a non monotonic change in spatial 
character of `near Fermi level' eigenstates with temperature, and the appearance 
of a weak pseudogap near $T_c$. 
\end{abstract}

\maketitle

Ferromagnetic polarons (FP) were proposed first in the context 
of Eu based magnetic semiconductors in the 1960s \cite{ref_other1,ref_EuS}.
Most of these materials display metallic resistivity,
$d\rho/dT > 0$, at low temperature, a strong peak around 
the ferromagnetic $T_c$, and a wide 
window with $d\rho/dT < 0$ 
thereafter \cite{ref_other1,ref_other_Eu_materials,ref_exp2,
ref_exp3,ref_exp_GdSi,ref_exp_GdN,ref_exp4}.
It was suggested 
\cite{ref_other_Eu_materials,ref_exp1,ref_exp2,ref_exp4,ref_exp_EuO}
that electrons in these low carrier density systems, 
while delocalised in the ferromagnetic
state, get `self trapped' in the paramagnetic state  
- creating a small polarised region and gaining energy by
localising there. Over the decades the transport in 
a large number of local moment ferromagnets has been 
attributed to FP 
\cite{ref_other_Eu_materials,ref_exp2,ref_exp3,ref_exp_GdSi,
ref_exp_GdN,ref_exp4}.

The intuitive physical picture has proved hard to convert to a 
concrete theory that addresses the accumulated data consistently.
The best documented material is EuB$_6$
\cite{ref_exp_opt_cond1,ref_exp_structure,ref_exp_sp_heat,
ref_exp_raman,ref_exp_field,ref_exp_carrier_density,ref_exp_muon,
ref_exp_magneto_optical,ref_exp_hall,ref_exp_opt_cond,ref_exp_resistivity,
ref_exp_rixs,ref_exp_strain,ref_exp_stm,ref_exp_stm_defects,
ref_exp_arps,ref_exp_topology,ref_exp_qnemetic}, 
a local moment ferromagnet with $S=7/2$  moments arising from Eu,
low carrier density, $n \sim 10^{-2}$ carriers
per formula unit, and a $T_c \sim 15$K. 
Raman scattering  
\cite{ref_exp_raman}
and muon spin relaxation 
\cite{ref_exp_muon}
had suggested  an inhomogeneous electronic-magnetic state 
near $T_c$.  The most unconventional features however 
are: (i)~The d.c resistivity, which shows a maximum near $T_c$, 
and then a shallow minimum at  $T \sim 2T_c$, with the
maximum-minimum feature suppressed by a field of $\sim 0.3$~Tesla
\cite{ref_exp_structure,ref_exp_hall}.
(ii)~Scanning tunneling spectroscopy (STS)
\cite{ref_exp_stm},
which reveals a spatial inhomogeneity in local conductance
near $T_c$ - with the textures having a linear dimension
$\sim 8$ lattice spacings and a showing 
tunneling peak at $\sim 25$mV below Fermi level.
These key experiments remain unexplained.

Several theoretical attempts have been made to model the 
magnetic polaron problem \cite{ref_theo_other_montecarlo,
ref_theory_1polaron,ref_theo_other_variation,ref_theo_other_dmft} and 
also specifically address the properties of EuB$_6$ 
\cite{ref_theory_polaron_hopping,
ref_theory_magneto_resistance,ref_theory_resistivity_phonon,ref_yu2005,
ref_theory_ising}. These include study of a 
polaron hopping model \cite{ref_theory_polaron_hopping} to address the 
$T \gg T_c$  resistivity, a small, $8 \times 8$, lattice Monte Carlo
to explain the specific heat \cite{ref_yu2005, ref_theory_ising} in terms 
of `polaron percolation', and transport models that invoke magnons 
and phonons \cite{ref_theory_resistivity_phonon}. However, none 
have explained the peculiar transport and spatial behaviour 
for reasonable parameter values.

The problem remains difficult because 
(i)~self trapping occurs in a finite temperature spatially
correlated spin background, and (ii)~the phenomena 
occurs at strong electron-spin coupling where we have no 
analytical tools for the electronic states.
This forces one to use numerical approaches and there the
difficulty is with accessible spatial scales. A single
polaron, at realistic electron-spin coupling, can have a
linear size $\sim 10$ lattice spacings and a 
finite electron number $N_{el}$, to mimic finite density, 
in a two dimensional (2D) system
will require $\sim N_{el} \times 10^2$ lattice sites. 
This has remained out of reach, so neither FP formation 
nor it's impact on magnetotransport and tunneling spectra 
have been addressed.  We offer a computational advance 
that allows access to these features.

In the standard model of FPs
\cite{ref_theo_other_montecarlo, ref_theory_1polaron, ref_theo_other_variation}
one has a ferromagnetic
Heisenberg model, with coupling $J$, with the spins
Kondo coupled to tight binding electrons (which have 
a hopping $t$) through an 
interaction $J'$. The magnetic states of this  
model can be accessed by a Monte Carlo (MC) scheme 
that iteratively diagonalises the electron problem 
\cite{ref_theo_other_montecarlo, ref_theory_1polaron, ref_yu2005, ref_theory_ising}.  
For a lattice with $N$ sites each microscopic update
costs ${\cal O}(N^3)$, and a `system sweep' costs
${\cal O}(N^4)$. This has limited studies to
sizes $\sim 10 \times 10$ 
\cite{ ref_yu2005, ref_theory_ising}. 
We set up a Langevin equation
(LE) to evolve the spins in the presence of a thermal noise
and sample configurations once equilibrium is reached.
The diagonalisation cost for torque calculation 
is still  ${\cal O}(N^3)$ but now
the spins can be updated in parallel, so the system
update cost is also ${\cal O}(N^3)$. This advantage 
allow a breakthrough, with 
access to lattices upto $20 \times 20$
and $N_{el} \sim 10$ (density $\sim 0.02$).

We use the LE approach on a 2D square lattice for
parameters that are generally accepted for EuB$_6$ 
to attempt as quantitative a match as possible. 
We set $t=100$meV, $J'S = t$, $JS^2 = 0.01t$,
\cite{ref_theory_polaron_hopping,ref_theory_magneto_resistance} and
electron density $n\sim 0.02$, probing the system as a 
function of temperature $T$ and magnetic field $h$.
Our main results are the following:

{\it 1.~Spatial textures in tunneling:}
Spatial textures in electron density and magnetisation, on the 
scale of $6-8$ lattice spacings, exist over a window $T \sim 
0.7T_c-1.5T_c$ at $h=0$. The local density of states 
(LDOS) in the `polaronic' 
regions have a distinct signature, with a peak at $\sim 25$ meV 
below Fermi level. This `binding energy' 
and the  spatial scale are in excellent agreement with experiment.  

{\it 2.~Resistivity, magnetotransport:}
The resistivity $\rho(T)$ is nonmonotonic, with a peak near $T_c$ 
and a shallow minimum at $T \sim 1.5 T_c$.  An applied field diminishes 
the peak-dip feature and suppresses it altogether by  $h \sim 1.0$ 
Tesla, consistent with EuB$_6$ data.

{\it 3.~Optical conductivity:}
We predict that the low frequency optical conductivity $\sigma(\omega,T)$ 
should have the expected Drude character for $T \ll T_c$ but by $0.7T_c$ 
it would pick up a distinct non Drude form, with the peak in $\sigma(\omega)$ 
shifting from $\omega=0$ to $\omega \sim 0.1t$ by $T_c$ and to $\omega 
\sim 0.2t$ by $2T_c$.

{\it 4.~Physical mechanism:}
The anomalies in local density of states and transport arise from 
(i)~the separation of the occupied (below chemical potential, $\mu$) 
states from the main body of the band by a weak 
pseudogap near $T_c$, and (ii)~a non monotonic degree of localisation 
of states in the window $\mu \pm k_BT$ 
as a function of temperature.

The Heisenberg-Kondo (H-K) model we study in 2D is:
\begin{equation}
H = 
 -t \sum_{\langle ij \rangle ,\sigma} 
c^{\dagger}_{i\sigma} c_{j\sigma}
- J'\sum_i {\bf S}_i.{\vec \sigma}_i 
-J \sum_{\langle ij \rangle} {\bf S}_i.{\bf S}_j
\end{equation}
The moments on Eu are large, with $2S \gg 1$ so we
treat them as classical unit vectors, absorbing 
the magnitude $S$ in the
coupling constants: $J \rightarrow JS^2$ and
$J' \rightarrow J'S$.
The parameter values have been stated earlier,
our electron number $N_{el} = 10$.

To generate equilibrium spin configurations
$\{{\bf S}_i \}$ we use a LE where
the spins experience a 
torque (below) derived from $H$ alongside 
a stochastic kick, with variance $\propto k_BT$, 
to model the effect of temperature.  This method 
not only facilitates the exploration of equilibrium 
spin configurations but also enables the study of
spin dynamics. The LE has the form:
\begin{eqnarray}
\frac{d\mathbf{S}_i}{dt} &=& \mathbf{S}_i \times (\mathbf{T}_i 
+ \mathbf{h}_i) - \gamma \mathbf{S}_i \times (\mathbf{S}_i 
\times \mathbf{T}_i) \cr
\mathbf{T}_i &=& -\frac{\partial H}{\partial \mathbf{S}_i} =
-J' \langle \vec{\sigma}_i \rangle -
J \sum_{\delta} {\bf S}_j \delta_{j,i+\delta}
\cr
\langle {h}_{i \alpha} \rangle &=& 0,~~ 
\langle
 h_{i \alpha}(t)h_{j \beta}(t') \rangle = 2\gamma k_B T
 \delta_{ij} \delta_{\alpha \beta} \delta(t-t') 
\end{eqnarray}
 $\mathbf{T}_i$ 
is the effective torque acting on the spin at the $i$-th 
 site, $\gamma =0.1 $
 is a damping constant, and $\mathbf{h}_i$ 
is the thermal noise satisfying the fluctuation-dissipation
 theorem.
$\langle \vec{\sigma}_i \rangle$ represents the 
expectation of $\vec{\sigma}_i$ taken over the
instantaneous ground state of the electrons, and
$\delta$ is the set of nearest neighbours. 
In the presence of an external magnetic field there is
an additional term ${\hat z} h$ in ${\bf T}_i$. 
Once Langevin evolution reaches equilibrium the magnetic
correlations can be computed from the 
$\{{\bf S}_i \}$, and electronic features obtained
by diagonalisation in these backgrounds (see Supplement).

\begin{figure}[b]
\centerline{
\includegraphics[height=4.2cm,width=6.5cm]{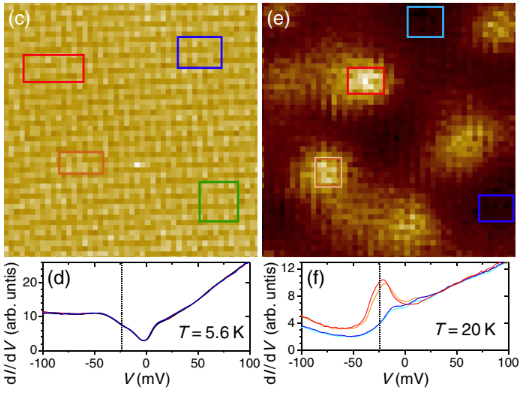}
}
\centerline{
~~~
\includegraphics[height=5.2cm,width=7.6cm]{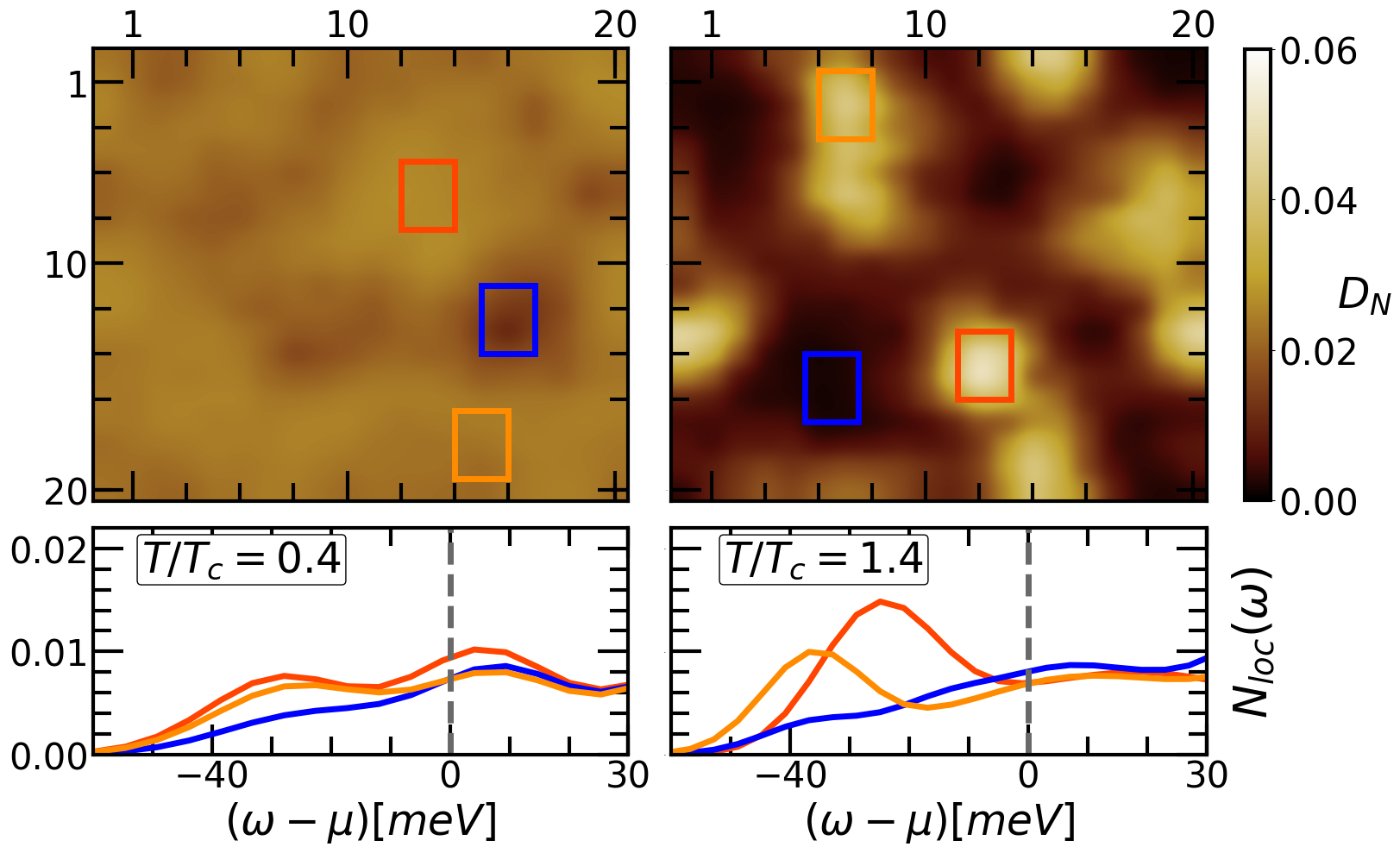}
}
\centerline{
~~~
\includegraphics[height=5.2cm,width=7.6cm]{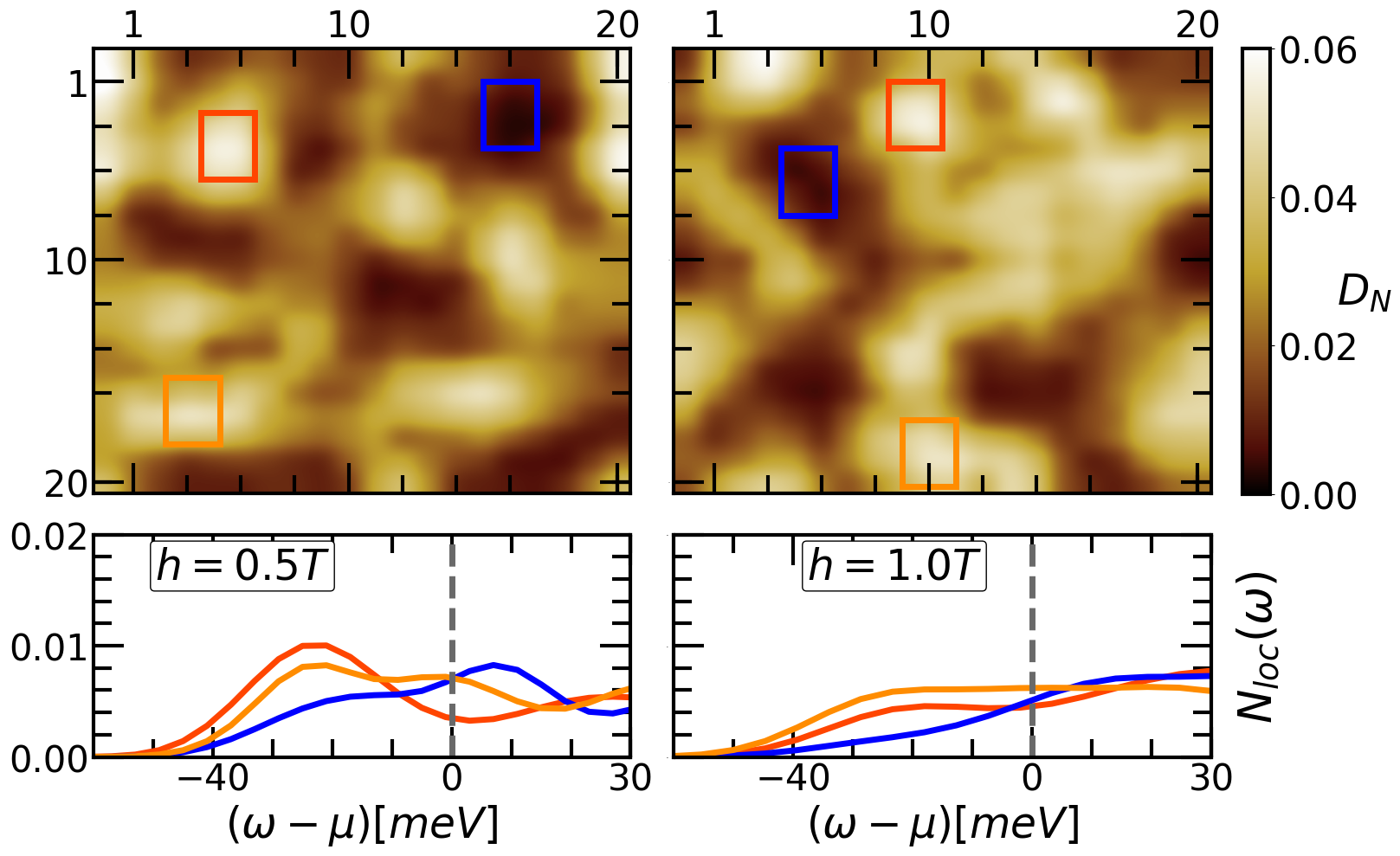}
}
\caption{Spatial map of tunneling conductance. Top row:
Experimental map of local conductance $g({\bf r},V)=dI({\bf r},V)/{dV}$
at $V=-24$ mV at $5.6$ K and $20$ K, with $T_c \sim 15K$. At
$T = 5.6$K the $-24$mV conductance is uniform. For $T
= 20$K it is inhomogeneous - the bright patches are
supposed to be the polaronic regions. The LDOS below shows that
$g(V)$ is site independent for $T \ll T_c$ and strongly
inhomogeneous for $T > T_c$.  Middle row: Our $h=0$ 
results on density $n({\bf r})$ at $T=0.4 T_c$ (left)
- showing a homogeneous density, and $T=1.4T_c$ (right) -
showing strong inhomogeneity. Corresponding LDOS panels
show almost site independent LDOS at $0.4T_c$ and a
strongly site differentiated LDOS at $1.4T_c$.
High $n({\bf r})$ regions have a peak at $\sim -25$meV.
Bottom row: $h$ dependence at $T=1.4T_c$. Left - $h=0.5$ Tesla,
right - $1.0$ Tesla. Trend: progressive homogenisation  
of $n({\bf r})$ and a site independent LDOS.
}
\end{figure}

{\it  Spatial textures in tunneling:}
Fig.1 reproduces spatial textures observed experimentally 
\cite{ref_exp_stm}
and compares them to our results.
The top row shows the experimental maps of local conductance
$g({\bf r},V)=dI({\bf r},V)/dV$ in EuB$_6$ at $V=-24$mV. We will
call the measured conductance at $V=-24$mV as $g_{ref}({\bf r})$.
The map on the left reveals homogeneous behaviour of  $g_{ref}$ 
at $T=5.6$ K ($\sim 0.4T_c$), which changes to a distinctly 
inhomogeneous behaviour at $T=20$ K ($ \sim 1.3T_c$) in
the right plot.  The linear dimension of the bright patches 
is $\sim 8$ lattice spacings. The panels below the spatial 
maps show $g({\bf r},V)$ averaged over the marked regions
in the upper panel. In the low $T$ case the $g(V)$ trace
does not depend on spatial location, while at $1.3T_c$ 
it shows very different $V < 0$ behaviour between the 
high $g_{ref}$ and low $g_{ref}$ areas, with a 
prominent peak at $V \sim -25$mV in the high 
$g_{ref}$ areas.

We directly calculate $n({\bf r})$, the electron density, using 
an instantaneous equilibrium spin configuration at a temperature 
$T$ as input and diagonalizing $H$. 
The spatial maps in the middle row show
$n({\bf r})$ at $T=0.4T_c$ (left) and $T=1.4T_c$ (right).
It is evident that at $T=0.4T_c$ carriers are
distributed almost uniformly across the lattice.
However, at  $T=1.4T_c$ the density is strongly
inhomogeneous.
Since $n({\bf r})  = \int_{-\infty}^{\infty} d\omega 
N({\bf r},\omega) f(\omega)$, where $N({\bf r},\omega)$ 
is the LDOS and $ f(\omega)$ is the Fermi function,
an inhomogeneity in $n({\bf r})$ implies a 
spatial variation of $N({\bf r},\omega)$.

Within the simplest approximation the tunneling conductance
at bias $V$ is proportional to the LDOS at $\omega = eV$,
where $e$ is the electron charge. We therefore plot 
our $N({\bf r},\omega)$ as a proxy for $g({\bf r},V)$.
As expected, at low $T$ the $ N({\bf r},\omega)$ traces 
are almost site indepedent. In the textured regime the
polaronic regions have a prominent peak at $\omega
\sim [-20, -30]$meV. The corresponding experimental
peaks are at $\sim -25$mV. The polaronic patches 
have size $\sim 8 \times 4$ lattice spacings.
The polaronic levels are centered roughly at 
$\sim 25$ meV below the Fermi energy and the
LDOS suggests a mild dip as we move into the unoccupied
part of the band.

\begin{figure}[b]
\centerline{
\includegraphics[height=7.8cm,width=8.2cm]{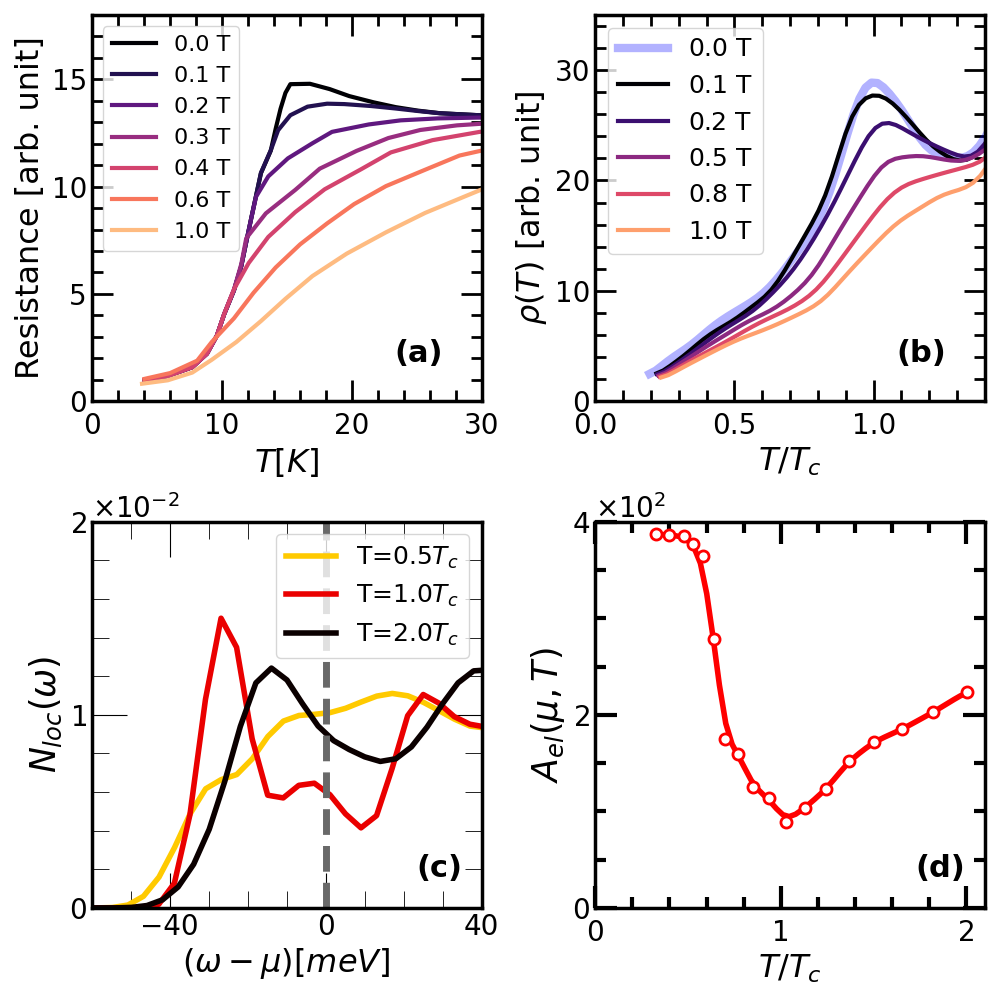}
~~
}
\caption{Resistivity. (a)~Temperature and field dependence 
of resistivity from experiments 
\cite{ref_exp_hall}. This is the total resistivity
without any subraction of a `phonon component' etc.
(b)~Our result showing the prominent peak feature at
$h=0$ and it's complete suppression at $h \sim 1$
Tesla.
(c)~The LDOS in polaronic regions as a function of $T$,
with a suppression near $\mu$ when $T \sim T_c$.
(d)~The area associated with near chemical potential
wavefunctions, averaged over $\epsilon \sim \mu \pm k_BT$,
as a function of $T$. High degree of localisation near $T_c$.
}
\end{figure}
\begin{figure}[t]
\centerline{
\includegraphics[height=5.0cm,width=8.3cm]{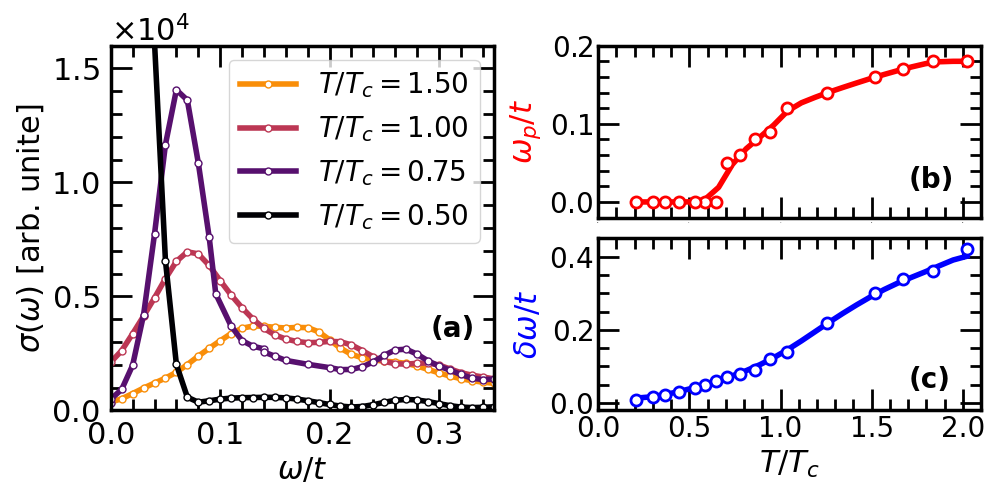}
~~
}
\caption{(a)~The low frequency behaviour of optical
conductivity is demonstrated for various temperatures.
The optical conductivity is plotted in measurable units
along the y-axis and frequency is plotted along the
x-axis scaled by hopping amplitude $t$. The
temperature, in units of $T_c$, is represented
through different colours in the plot.
(b)~$T$ dependence of $\omega_p$, the peak location of 
$\sigma(\omega)$. Finite $\omega_p$ indicates non Drude
behaviour.
(c)~$T$ dependence of of $\delta \omega$, the full width 
at half maximum of $\sigma(\omega)$.
}
\end{figure}

The bottom panels show the effect of an applied
magnetic field $h$ at $T=T_c$.
At $h=0.5$ Tesla (left) the white regions are already more 
spread out than at $h=0$ and the corresponding LDOS
has less prominent peaks than at $h=0$. At $h=1.0$ Tesla
(right) the polarons have `overlapped' and essentially 
occupy the whole system. The density contrast is low,
and the LDOS at the high density sites is only
slightly different from that at the low density sites. 
We will see the impact in the transport properties next.

{\it Resistivity and magnetoresistance:}
Unlike the monotonic $T$ dependence  
in conventional ferromagnetic metals the resistivity $\rho(T)$ 
in EuB$_6$ has a peak at $T \sim T_c$ a minimum near $2T_c$, 
and then rises again. The measured $\rho(T,h)$  is shown in 
Fig.2(a) for $T \le 2T_c$. The application of a magnetic field 
reduces the peak-dip feature, and by $h \sim 0.3$ Tesla it 
completely disappears. The experimental resistivity
has a significant phonon contribution, with $d\rho/dT > 0$,
which can be comparable to the magnetic scattering for
$ T \gg T_c$
\cite{ref_exp_field}, so
we compare 
our result, Fig.2(b), to the experiment only 
for $T \lesssim 1.5T_c$.
Our $h=0$ resistivity has a `peak-dip' 
feature, with a peak around $T_c$ and a minimum
at $T \sim 1.4T_c$, beyond which it rises again.
$\rho(T)$ in the window $h=0.1-1.0$ Tesla shows 
a clear suppression of the $d\rho/dT < 0$ behaviour 
with increasing $h$.  By $h=1.0$ Tesla the 
non-monotonicity is gone.

Non monotonic temperature dependence of resistivity,
with a $d\rho/dT < 0$ window, can arise either from 
the presence of a gap, or localisation of states near 
the chemical potential, $\mu$.
Our global DOS does not show any gap in the relevant
$T$ window. However for $T \sim 0.7T_c - 1.5T_c$ 
the polaronic regions show a LDOS, $N_{loc}(\omega)$, 
with enhancement for $\omega < \mu$ and a weak 
suppression at $\omega \sim \mu$, Fig.2(c). 
This weak `pseudogap' is absent both at 
$T \ll T_c$ and $T \gg T_c$. In fact at larger
coupling,  $J'=5t$, and for a single polaron, an
actual `gap' has been established before
\cite{ref_theory_1polaron}.
We also calculated $A_{el}$, the average 
spatial coverage of eigenstates 
in the energy window $\epsilon \in (\mu \pm k_BT)$
(see Supplement). 
For $T \ll T_c$ this is just system size, $L^2$, but
for $T \sim T_c$ this is only ${\cal O}(9^2)$,
i.e, polaron size. It
rises again as $T$ grows beyond $T_c$, Fig.2(d).
These observations 
do not constitute a `theory' for $\rho(T)$ near $T_c$ 
but correlate non monotonicity
in two simpler quantities with that observed in
$\rho(T)$.

{\it Optical conductivity:}
The temperature and field dependence of the resistivity correlate
with the presence of electronic-magnetic textures in the system.
We wanted to check whether the unusual electronic state had
a signature in the optical conductivity $\sigma(\omega)$ in
the polaronic window. 
Most of the available data focus on high energies  
\cite{ref_exp_opt_cond1, ref_exp_sp_heat, ref_exp_opt_cond},
where we
think the effect of a low energy bound state like the polaron 
would be limited. Fig.3(a) shows our result over a window
$\omega = 0-0.3t$, which would translate to $\sim 0-30$meV 
for EuB$_6$. 
We show this for four temperatures between $0.5T_c$ and
$1.5T_c$.
At $T = 0.5T_c$ where the magnetic state is highly
polarised and the electronic
state is essentially homogeneous we see a Drude response
with a sharp peak at $\omega=0$ of width $\sim 0.02t$.
The Drude feature sharpens as $T \rightarrow 0$. However
as $T$ increases we see that already by $0.7T_c$ the peak
location $\omega_p$ has shifted to a finite value and
the `d.c conductivity' is strongly suppressed. 
This non Drude response persists to high $T$. In
Fig.3(b ) we show 
$\omega_p(T)$ and in Fig.3(c) we show 
the width $\delta \omega(T)$.

\begin{figure}[t]
\centerline{
\includegraphics[height=5.5cm,width=3.9cm]{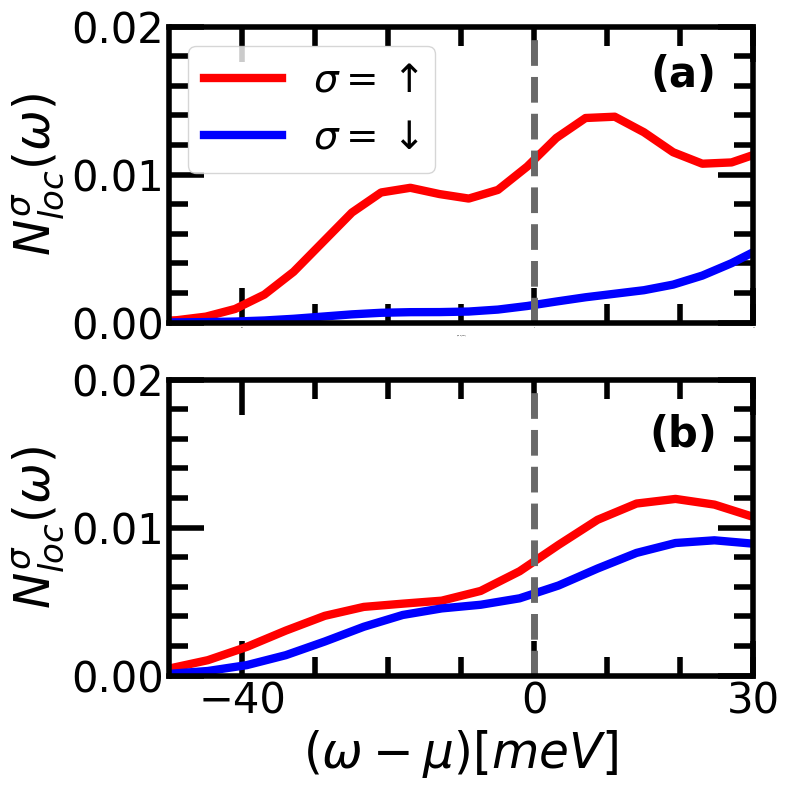}
\includegraphics[height=5.6cm,width=4.7cm]{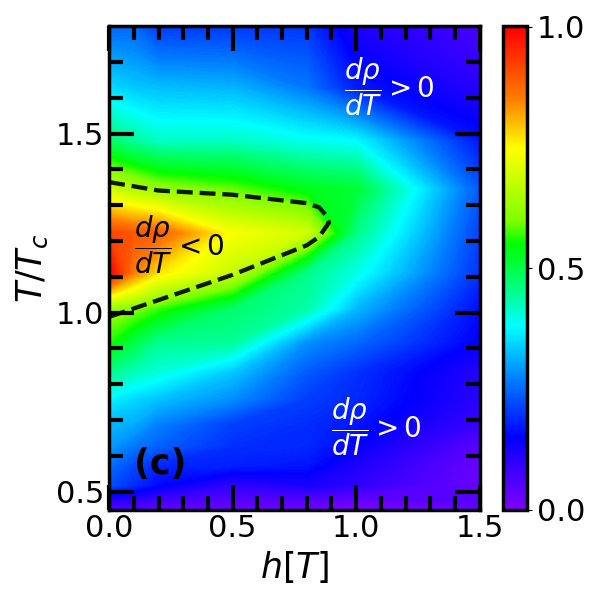}
}
\caption{Spin resolved LDOS and $h-T$ phase diagram.
(a)~SRLDOS for $h=0.01$T and $T=T_c$ in the
the polaronic region, and (b)~same for the non
polaronic spatial regions.
The red line represents the up spin component of the SRLDOS 
while the blue line represents the down spin component. 
(c)~Map of of the difference in integrated weight of up and
down spin spectra taken over the spatially inhomogeneous 
regions, for varying $h$ and $T$.
The difference $\delta W$ is maximum near $T_c$ at $h=0$
and falls off either as $T$ moves away or $h$ is increased.
There is no sharp boundary in terms of this quantity.  
The contour corresponding to $d \rho/dT =0$ (enclosing the
$ d \rho/dT < 0$ regime) delineates prominent
polaronic behaviour.
}
\end{figure}

{\it $h-T$ phase diagram:}
We want to combine the results obtained on the various
measurables to create a `polaronic phase diagram'
for EuB$_6$. While scanning tunneling (or LDOS) 
measurement is the most direct indicator of
density inhomogeneity, by itself it does not reveal
whether there is a large `polarisation' in the
high density region. A minor variation, probing
the spin resolved LDOS (SRLDOS) in a small
field, would not only indicate
regions of high total density (by summing the up and
down components) but also the extent of polarisation
(from the difference of up and down components). 
Fig.4(a) shows the site-averaged SRLDOS
for polaronic sites while Fig.4(b) shows it for
for the non polaronic sites, both at $h=0.01$T and $T=T_c$.
The up-down difference in 4(a), versus the
up-down similarity in 4(b), is readily
visible.

We compute 
 $\delta W=|W_{\uparrow} - W_{\downarrow}|$, where   
 $W_{\uparrow}= -(1/\pi) \frac{1}{N_p}\sum_i
 \int_{-\infty}^{\mu} Im[N_{loc\ ii}^{\uparrow}(\omega)]
  d\omega$, etc,
where $N_p$
 is the number of sites in the polaronic region, and the sum  
 runs over these $N_p$ sites. 
We plot a normalised $\delta W$ (dividing by it's maximum
value for any $h, T$) with respect to $h$ and $T$ in 
Fig.4(c).  There is no phase transition between the polaronic 
and non polaronic regimes, it is a crossover.  
There is a specific line one can draw, bounding the
$d\rho/dT < 0$ window and that is shown by the dashed line.

Finally, We discuss a few issues which have bearing on the 
theory-experiment comparison. (i)~Effect of changing $J'$: we have 
checked that polarons are not stable below $J' \sim 0.5t$. 
For $0.5t < J < 2t$ there is a peak-dip
feature in $\rho(T)$. The peak-dip height difference increases with
$J'$ but the associated temperature window  remains 
roughly $J'$ independent.
(ii)~Effect of impurities: we have not done a full calculation
including impurities but anticipate that low enough disorder (in 3D)
will leave the low $T$ behaviour intact but can enhance the $d \rho/dT
< 0$ window in temperature. It will also introduce pinning centers for
the polarons.
(iii)~Effect of dimensionality: the physical process that leads to
FP formation is dimension independent but some of the pathologies
related to 2D, e.g, disorder induced localisation, or lack of
long range order at finite $T$, will be absent.
We will present some results separately.
(iv)~Several materials consisting of rare earth elements, such as 
EuS, EuO, and GdN, exhibit ferromagnetic ground states and 
low magnetic transition temperatures (12K-70K), and show
nonmonotonic resistivity.
Typically, the resistivity peak occurs around
$T_c$, but the peak-to-dip ratios differ. 
Although the variation in
the peak-to-dip ratio can be captured by 
varying $J'/t$ and carrier density the 
strongly insulating high $T$ behaviour in some materials, 
we think, requires the inclusion of pinning centers.

{\it Conclusions:}
We resolve the longstanding problem of transport and spectral 
features in EuB$_6$, the candidate material for ferromagnetic 
polarons, by using an exact diagonalisation based
Langevin approach for the thermal state of low density 
electrons coupled to a Heisenberg spin system.
We use intermediate electron-spin coupling, as suggested
by experiments, and find a non monotonic resistivity with
a small peak near $T_c$. Our spatially
resolved tunneling density of states shows a striking
inhomogeneity in a temperature window near $T_c$, with
spatial scale and spectral features that are in excellent
agreement with recent experiments. The resistivity
and tunneling features are related to a 
strongly temperature
dependent pseudogap in the electronic spectrum
and the localisation of near Fermi level states. 
We make two specific predictions: 
(i)~the optical conductivity 
would evolve from a Drude response at $T \ll T_c$ to         
having a finite frequency peak, with characteristic
location and width, as $T \rightarrow T_c$, and 
(ii)~polaronic signatures, indicated by the 
spin resolved local density of states, 
would be limited in EuB$_6$ to 
$T \sim 0.7T_c-1.5T_c$ and $h \sim 0-1$ Tesla.
Our approach, augmented by the presence of pinning
centers,
can address supposed polaronic signatures in a host
of rare earth materials.

 We acknowledge use of the HPC facility at HRI.

\vspace{.5cm}

\centerline{\bf Supplementary Material}

\vspace{.4cm}

\noindent
{\bf Computation of electronic properties}
\\

Let the equilibrium spin configurations generated by the
Langevin equation at some temperature $T$ be indexed by
a label $\alpha$, i.e, the spin configurations are
$\{{\bf S}_i \}_{\alpha}$.
The corresponding set of single particle eigenvalues would
be $\epsilon_n^{\alpha}$ and the eigenstates would be
$\psi^{\alpha}_n({\bf r})$.
The electron spin is not a quantum number in 
a generic ${\bf S}_i$ configuration so the index $n$
has $2N$ values where $N$ is the number of lattice sites.
In what follows $f(\epsilon) = (e^{{\beta}(\epsilon - \mu)} + 1)^{-1}$
is the Fermi function.

We first write the expressions for various electronic
properties in a specific configuration $\alpha$. 
Since the system is translation invariant, i.e, without
any extrinsic disorder, quantities like local density and
LDOS will be site independent on configuration averaging.
For them the results we show pertain to single configurations.
For the optical conductivity and resistivity, however,
configuration average is meaningful and we do so over
typically $100$ configurations.
\begin{enumerate}
\item
Spatial density: 
\begin{equation}
n_{\alpha}({\bf r}) 
= \sum_n \vert \psi_n^{\alpha}({\bf r}) \vert^2 f(\epsilon_n^{\alpha})
\end{equation}
\item
Density of states.
\begin{equation}
N_{\alpha}(\omega) = \sum_n \delta(\omega - \epsilon_n^{\alpha})
\end{equation}
\item
Local density of states.
\begin{equation}
N_{\alpha}({\bf r}, \omega) = \sum_n 
\vert \psi_n^{\alpha}({\bf r}) \vert^2  \delta(\omega - \epsilon_n^{\alpha})
\end{equation}
\item
Spin resolved DOS: the LDOS above is spin summed. To know if there is a large
local magnetisation over a spatial neighbourhood it is useful to calculate
the spin resolved LDOS from the local Greens 
function $G^{\alpha}_{\sigma \sigma}({\bf r}, \omega)$.
\begin{eqnarray}
N^{\alpha}_{\sigma}({\bf r}, \omega) & = & - ({1 \over \pi}) Im 
[G^{\alpha}_{\sigma \sigma}({\bf r}, \omega)] 
\cr
\cr
G^{\alpha}_{\sigma \sigma}({\bf r}, \omega) &=&
\int_0^{\infty} 
dt e^{i \omega t} \langle \Psi^0_{\alpha} \vert 
[c_{{\bf r} \sigma}(t), c^{\dagger}_{{\bf r} \sigma}(0)]
\vert \Psi^0_{\alpha} \rangle
\end{eqnarray}
where $\vert \Psi^0_{\alpha} \rangle$ is the many body ground state
of the $N_{el}$ system (here we ignore Fermi factors).
\item
Optical conductivity.
\begin{eqnarray}
\sigma_{\alpha} (\omega) &=& A
\sum_{m,n} 
{ {f(\epsilon_m^{\alpha}) - f(\epsilon_n^{\alpha})}
\over {\epsilon_m^{\alpha} - \epsilon_n^{\alpha}} }
\vert {\hat j}_{mn}^{\alpha \alpha} \vert^2
\delta(\omega - ({\epsilon_m^{\alpha} - \epsilon_n^{\alpha}}))
\cr
{\hat j} &=& i t a_0 e \sum_{i, \sigma} (c^{\dagger}_{{i + x a_0},\sigma}
c_{i, \sigma} - h.c) 
\end{eqnarray}
where $A = {\pi e^2}/{\hbar N}$
\item
DC resistivity:
we obtained the `dc conductivity' $\sigma_{\alpha}$ as
\begin{equation}
\sigma_{\alpha} = {1 \over {\Delta \omega}} 
\int_0^{\Delta \omega} d \omega \sigma_{\alpha} (\omega)
\end{equation}
where $\Delta \omega$ is a small multiple of the average finite
size gap in the spectrum. In our calculations it is $0.05t$.
The d.c resistivity is the inverse of the thermally averaged
conductivity
$$
\rho = {1 \over {N_{\alpha}}} \sum_{\alpha} {1 \over {\sigma_{\alpha}}}
$$
\item
Inverse participation ratio (IPR): this is useful to quantify the
inverse of the
`volume' associated with a single particle eigenstates. The standard
definition, for normalised states, is:
\begin{equation}
I^{\alpha}_n = \int d{\bf r} \vert \psi_n^{\alpha}({\bf r}) \vert^4
\end{equation}
D.c transport involves states over a window $\mu \pm k_BT$. Since the
character of states changes rapidly with energy in the polaronic
regime we calculate the typical area $A(T)$ associated with eigenstates
in the  $\mu \pm k_BT$ window as follows:
\begin{equation}
A_{\alpha}(T) = {1 \over N_S} \sum_n {1 \over {I^{\alpha}_n}}~,
~~~~~{\epsilon_n \in (\mu \pm k_BT)} 
\end{equation}
where $N_S$ is the number of states in the $\mu \pm k_BT$ window.
We then average $A_{\alpha}$ over configurations.
\end{enumerate}

\end{document}